\documentclass{revtex4}
\newdimen\SaveWidth \SaveWidth=\textwidth
\newdimen\SaveHeight \SaveHeight=\textheight
\advance\SaveWidth by -\textwidth
\advance\SaveHeight by -\textheight
\divide\SaveWidth by 2
\divide\SaveHeight by 2
\advance\hoffset by \SaveWidth
\advance\voffset by \SaveHeight
\def\abs#1{\left| #1\right|}

\def\abs#1{\left| #1\right|}

\def\etmiss{\slashchar{E}_T}

\def\tell{{\tilde\ell}}
\def\ttau{{\tilde\tau}}
\def\fbi{{\rm fb}^{-1}}
\def\Meff{M_{\rm eff}}

\def\lsp{{\tilde\chi_1^0}}

\def\GeV{{\rm GeV}}
\def\TeV{{\rm TeV}}

\def\tchi{\tilde\chi}
\def\tg{\tilde g}
\def\tq{\tilde q}

\let\badcite=\cite
\def\cite{~\badcite}

\def\cmsec{{\rm cm}^{-2}{\rm s}^{-1}}

\def\jet{{\rm jet}}

\def\slashchar#1{\setbox0=\hbox{$#1$}           
   \dimen0=\wd0                                 
   \setbox1=\hbox{/} \dimen1=\wd1               
   \ifdim\dimen0>\dimen1                        
      \rlap{\hbox to \dimen0{\hfil/\hfil}}      
      #1                                        
   \else                                        
      \rlap{\hbox to \dimen1{\hfil$#1$\hfil}}   
      /                                         
   \fi}                                         %

\catcode`@=11
\newdimen\vbigd@men                             

\def\vbig#1#2{{\vbigd@men=#2\divide\vbigd@men by 2%
   \hbox{$\left#1\vbox to \vbigd@men{}\right.\n@space$}}}
\catcode`@=12

\def\simge{
    \mathrel{\rlap{\raise 0.511ex
        \hbox{$>$}}{\lower 0.511ex \hbox{$\sim$}}}}
\def\simle{
    \mathrel{\rlap{\raise 0.511ex 
        \hbox{$<$}}{\lower 0.511ex \hbox{$\sim$}}}}

\catcode`@=11
\def\citenum#1{\csname b@#1\endcsname}
\catcode`@=12

\usepackage{graphicx}
\begin{document}
\bibliographystyle{revtex}


\title{
High-Mass Supersymmetry with High Energy Hadron Colliders}



\author{I. Hinchliffe}
\email[]{I_Hinchliffe@lbl.gov}
\thanks{The work was supported in part by the Director, Office of Energy
Research, Office of High Energy and Nuclear Physics of the
U.S. Department 
of Energy under Contract
DE--AC03--76SF00098.}
\affiliation{Lawrence Berkeley National Laboratory, Berkeley, CA}
\author{F.E. Paige}
\email[]{paige@bnl.gov}
\thanks{The work was supported in part by the Director, Office of Energy
Research, Office of High Energy and Nuclear Physics of the
U.S. Department
 of Energy under Contract
DE-AC02-98CH10886.}
\affiliation{ Brookhaven National Laboratory, Upton, NY}


\date{\today}

\begin{abstract}

While it is natural for supersymmetric particles to be well within the
 mass range of the large hadron collider, it is possible that the
 sparticle masses could be very heavy.
Signatures are examined at a very high energy hadron collider and an
very high luminosity option for the Large Hadron Collider 
 in such scenarios.
\end{abstract}

\maketitle

\section{Introduction}

If supersymmetry is connected to the hierarchy problem, it is
expected \cite{fine-tuning} that sparticles will be sufficiently light that at least
some of them will be observable at the Large Hadron Collder (LHC) or
even at the Tevatron. 
However it is not possible to set a rigorous bound on the sparticle
masses.  As the
sparticle masses rise, the fine tuning problem of the standard model
reappears and  the sparticle masses become large enough so that they are
difficult to observe at LHC.

\begin{table}[t]
\caption{Benchmark SUGRA points and masses from Ref.~\cite{Battaglia:2001zp}
\label{points}}
\medskip
\renewcommand{\arraystretch}{0.95}
\begin{centering}
\small
\begin{tabular}{|c||r|r|r|r|r|r|r|r|r|r|r|r|r|}
\hline
Model          & A   &  B  &  C  &  D  &  E  &  F  &  G  &  H  &  I  &  J  &  K  &  L  &  M   \\ 
\hline
$m_{1/2}$      & 600 & 250 & 400 & 525 &  300& 1000& 375 & 1500& 350 & 750 & 1150& 450 & 1900 \\
$m_0$          & 140 & 100 &  90 & 125 & 1500& 3450& 120 & 419 & 180 & 300 & 1000& 350 & 1500 \\
$\tan{\beta}$  & 5   & 10  & 10  & 10  & 10  & 10  & 20  & 20  & 35  & 35  & 35  & 50  & 50   \\
sign($\mu$)    & $+$ & $+$ & $+$ & $-$ & $+$ & $+$ & $+$ & $+$ & $+$ & $+$ & $-$ & $+$ & $+$  \\ 
$\alpha_s(m_Z)$& 120 & 123 & 121 & 121 & 123 & 120 & 122 & 117 & 122 & 119 & 117 & 121 & 116 \\
$m_t$          & 175 & 175 & 175 & 175 & 171 & 171 & 175 & 175 & 175 & 175 & 175 & 175 & 175  \\ \hline
Masses         &     &     &     &     &     &     &     &     &     &     &     &     &      \\ \hline
h$^0$          & 114 & 112 & 115 & 115 & 112 & 115 & 116 & 121 & 116 & 120 & 118 & 118 &  123 \\
H$^0$          & 884 & 382 & 577 & 737 &1509 &3495 & 520 &1794 & 449 & 876 &1071 & 491 & 1732 \\
A$^0$          & 883 & 381 & 576 & 736 &1509 &3495 & 520 &1794 & 449 & 876 &1071 & 491 & 1732 \\
H$^{\pm}$      & 887 & 389 & 582 & 741 &1511 &3496 & 526 &1796 & 457 & 880 &1075 & 499 & 1734 \\ \hline
$\chi^0_1$     & 252 &  98 & 164 & 221 & 119 & 434 & 153 & 664 & 143 & 321 & 506 & 188 &  855 \\
$\chi^0_2$     & 482 & 182 & 310 & 425 & 199 & 546 & 291 &1274 & 271 & 617 & 976 & 360 & 1648 \\
$\chi^0_3$     & 759 & 345 & 517 & 654 & 255 & 548 & 486 &1585 & 462 & 890 &1270 & 585 & 2032 \\
$\chi^0_4$     & 774 & 364 & 533 & 661 & 318 & 887 & 501 &1595 & 476 & 900 &1278 & 597 & 2036 \\
$\chi^{\pm}_1$ & 482 & 181 & 310 & 425 & 194 & 537 & 291 &1274 & 271 & 617 & 976 & 360 & 1648 \\
$\chi^{\pm}_2$ & 774 & 365 & 533 & 663 & 318 & 888 & 502 &1596 & 478 & 901 &1279 & 598 & 2036 \\ \hline
$\tilde{g}$    &1299 & 582 & 893 &1148 & 697 &2108 & 843 &3026 & 792 &1593 &2363 & 994 & 3768 \\ \hline
$e_L$, $\mu_L$ & 431 & 204 & 290 & 379 &1514 &3512 & 286 &1077 & 302 & 587 &1257 & 466 & 1949 \\
$e_R$, $\mu_R$ & 271 & 145 & 182 & 239 &1505 &3471 & 192 & 705 & 228 & 415 &1091 & 392 & 1661 \\
$\nu_e$, $\nu_{\mu}$
               & 424 & 188 & 279 & 371 &1512 &3511 & 275 &1074 & 292 & 582 &1255 & 459 & 1947 \\
$\tau_1$       & 269 & 137 & 175 & 233 &1492 &3443 & 166 & 664 & 159 & 334 & 951 & 242 & 1198 \\
$\tau_2$       & 431 & 208 & 292 & 380 &1508 &3498 & 292 &1067 & 313 & 579 &1206 & 447 & 1778 \\
$\nu_{\tau}$   & 424 & 187 & 279 & 370 &1506 &3497 & 271 &1062 & 280 & 561 &1199 & 417 & 1772 \\ \hline
$u_L$, $c_L$   &1199 & 547 & 828 &1061 &1615 &3906 & 787 &2771 & 752 &1486 &2360 & 978 & 3703 \\
$u_R$, $c_R$   &1148 & 528 & 797 &1019 &1606 &3864 & 757 &2637 & 724 &1422 &2267 & 943 & 3544 \\
$d_L$, $s_L$   &1202 & 553 & 832 &1064 &1617 &3906 & 791 &2772 & 756 &1488 &2361 & 981 & 3704 \\
$d_R$, $s_R$   &1141 & 527 & 793 &1014 &1606 &3858 & 754 &2617 & 721 &1413 &2254 & 939 & 3521 \\
$t_1$          & 893 & 392 & 612 & 804 &1029 &2574 & 582 &2117 & 550 &1122 &1739 & 714 & 2742 \\
$t_2$          &1141 & 571 & 813 &1010 &1363 &3326 & 771 &2545 & 728 &1363 &2017 & 894 & 3196 \\ 
$b_1$          &1098 & 501 & 759 & 973 &1354 &3319 & 711 &2522 & 656 &1316 &1960 & 821 & 3156 \\
$b_2$          &1141 & 528 & 792 &1009 &1594 &3832 & 750 &2580 & 708 &1368 &2026 & 887 & 3216 \\ 
\hline
\end{tabular}
\end{centering}
\end{table}

It is also possible that SUSY is also the solution to the dark matter
problem \cite{BB},  the stable,
lightest supersymmetric particle (LSP) being the particle that
pervades the universe. This constraint can be applied to the minimal
SUGRA \cite{sugra} model and used to constrain the masses of the other
sparticles.
Recently sets of paramters in  the minimal SUGRA model have been proposed~\cite{Battaglia:2001zp}
that satisfy existing constraints, including the dark matter
constraint and the one from the observed anomaly in the magnetic
moment of the muon~\cite{g-2},  but do not impose any fine tuning
requirements. This set of points is not a random sampling of the  available
parameter space but is rather intended to illustrate the possible
experimental consequences. These points and their mass spectra are shown in
Table~\ref{points}. Most of the allowed parameter space corresponds to
the case where the sparticles have masses less than 1 TeV or so and
is accessible to LHC. Indeed 
some of these points are quite similar to ones studied in earlier LHC
simulations \cite{cmssusy}\cite{AtlasPhysTDR}. 
Points A, B, C, D, E, G, J and L fall into this category.
As the masses of the sparticles are increased, the LSP contribution to
dark matter rises and typically violates the experimental
constraints. However there are certain regions of parameter space
where the annihilation rates for the LSP can be increased and the
relic density of LSP's lowered sufficiently. In these
narrow regions, the sparticle masses can be much larger. Points F, K, H
and M illustrate these regions. This paper 
 considers Point K, H and M 
at the LHC with a luminosity upgrade to $1000\,\fbi$
per year, (SLHC)
and at a possible higher energy hadron collider (VLHC). 
We assume an energy of $40\,\TeV$ for the VLHC
and use the identical analysis for both machines. 
 Point F has  similar pheonomenology to Point K except that the
squark and slepton masses are much larger and consequently more
difficult to observe.
For the purposes of this simulation, the detector
performance at $10^{35}\,\cmsec$  and at the VLHC 
is assumed to be the same as that of
ATLAS  for at the LHC design 
luminosity. In particular, the additional pileup present at higher
luminosity 
is taken into account only by raising some of the
cuts.  Isajet~7.54\cite{ISAJET}  is used for the event
generation. Backgrounds from $t\overline{t}$, gauge boson pairs, large
$p_T$ gauge boson production and QCD jets are included.

\section{Point K}

\begin{figure}
\includegraphics[width=3.0in]{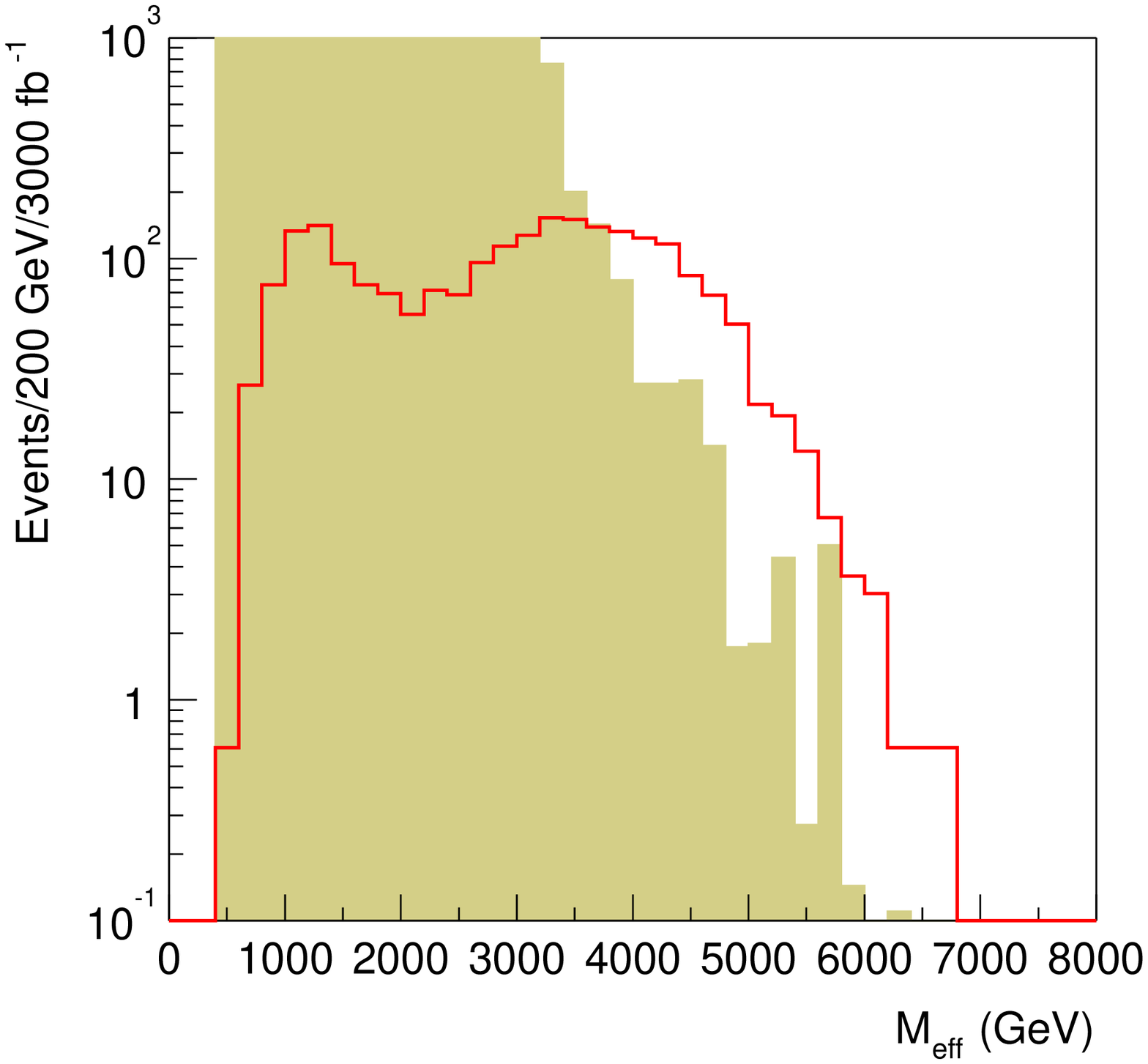}
\includegraphics[width=3.0in]{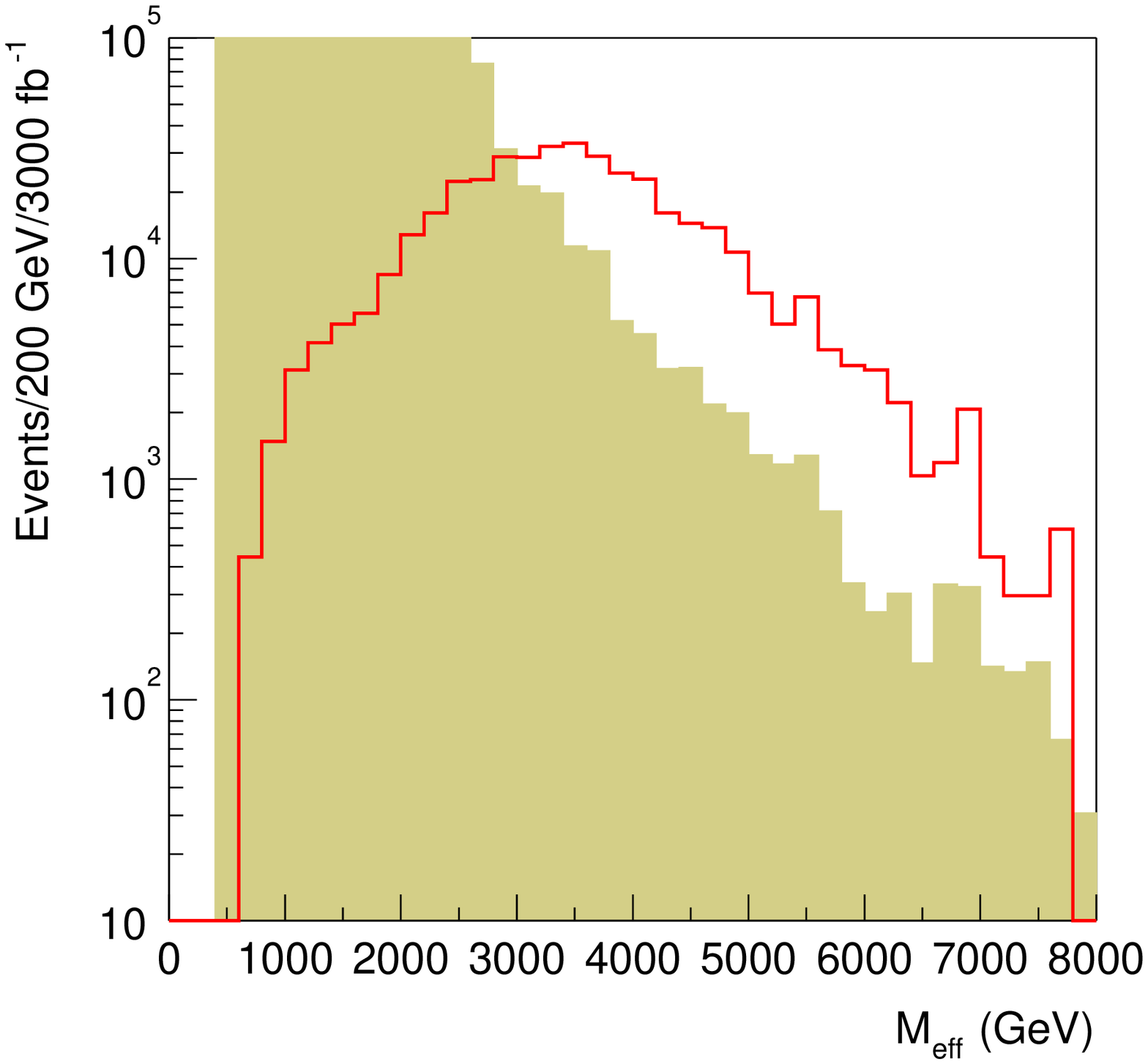}
\caption{$\Meff$ distribution for SLHC (left) and VLHC (right) for Point
K. Solid: signal. Shaded: SM background. \label{pointkmeff}}
\end{figure}

Point K has $M_A \approx 2M_\lsp$ and gluino and squark masses above
$2\,\TeV$. The strong production is dominated by valance squarks,
which have the characteristic decays
$\tq_L \to \tchi_1^\pm q, \tchi_2^0 q$ and $\tilde q_R \to \lsp q$.
The signal can be observed in the inclusive effective mass
distribution. Events are selected with hadronic jets and missing $E_T$
and the following scalar quantity formed
$$ M_{eff} = \etmiss + \sum_{jets}E_{T,jet} +\sum_{leptons}E_{T,lepton} $$
where the sum runs over all jets with $E_T> 50 $ GeV and
$\abs{\eta}<5.0$ and isolated leptons with  $E_T> 15 $ GeV and$\abs{\eta}<2.5$ . 
The following further selection was then made: events were selected 
with at least two jets
with $p_T > 0.1\Meff$, $\etmiss > 0.3\Meff$, $\Delta\phi(j_0,\etmiss) <
\pi-0.2$, and $\Delta\phi(j_0,j_1)<2\pi/3$. These cuts help to
optimize the signal to background ratio.
The distributions in  $ M_{eff}$ for signal and
background are shown in Figure~\ref{pointkmeff}. It can be seen that the
signal emerges from the background at large values of $M_{eff}$.
 The LHC  with 3000 $\fbi$ of integrated luminosity has a
signal of 510 events on a background of 108 for
$\Meff>4000\,\GeV$. These rates are sufficiently large so that a
discovery could be made with the standard integrated luminosity of 300
$\fbi$. However the limited data samples available will restrict
detailed studies.

\begin{figure}[t]
\includegraphics[width=3.0in]{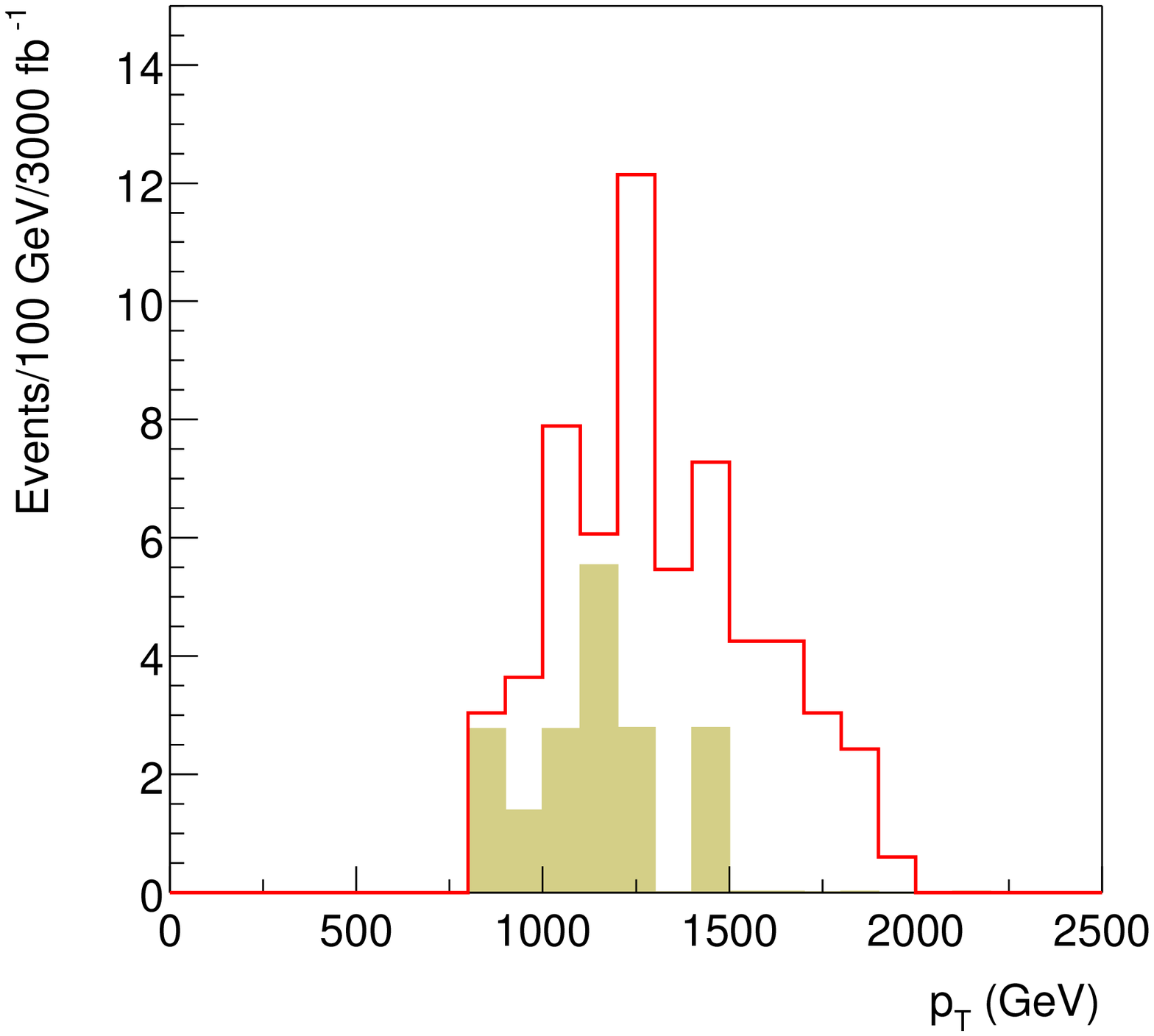}
\includegraphics[width=3.0in]{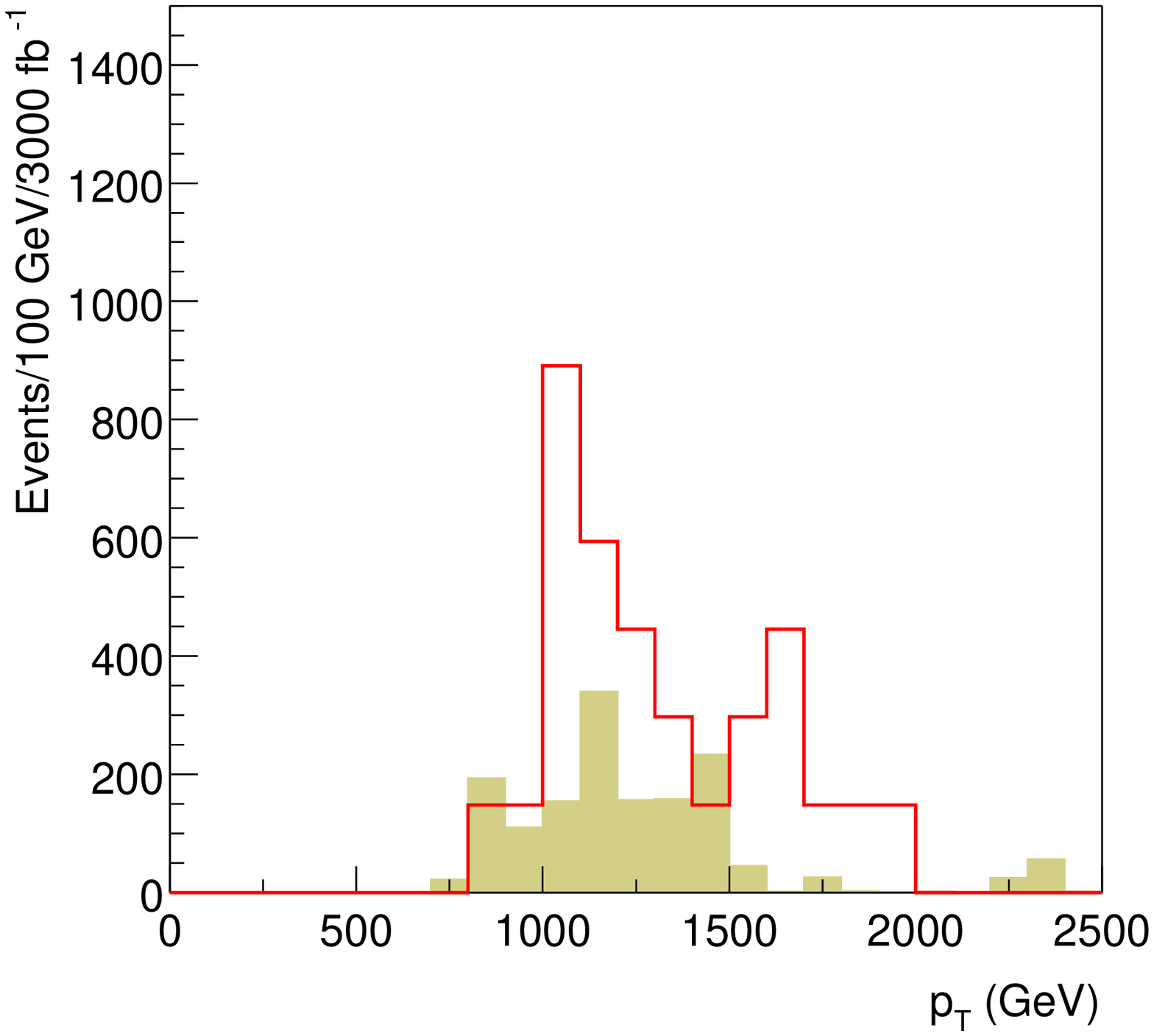}
\caption{$p_T$ distribution of hardest jet in $2\jet+\etmiss$ events for
SLHC (left) and VLHC (right) for Point K. \label{pointkptfroid}}
\end{figure}
\begin{figure}
\includegraphics[width=3.0in]{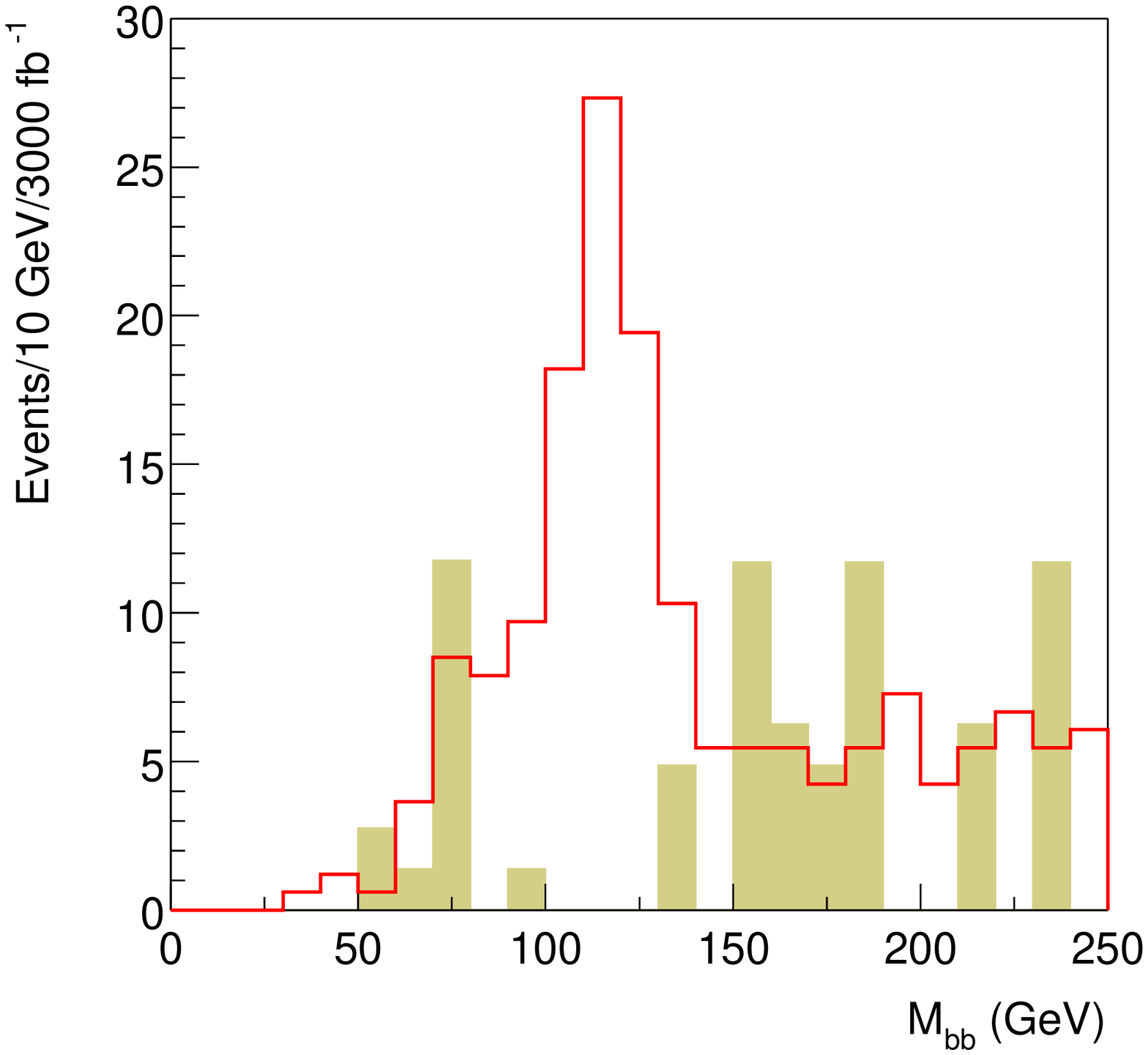}
\includegraphics[width=3.0in]{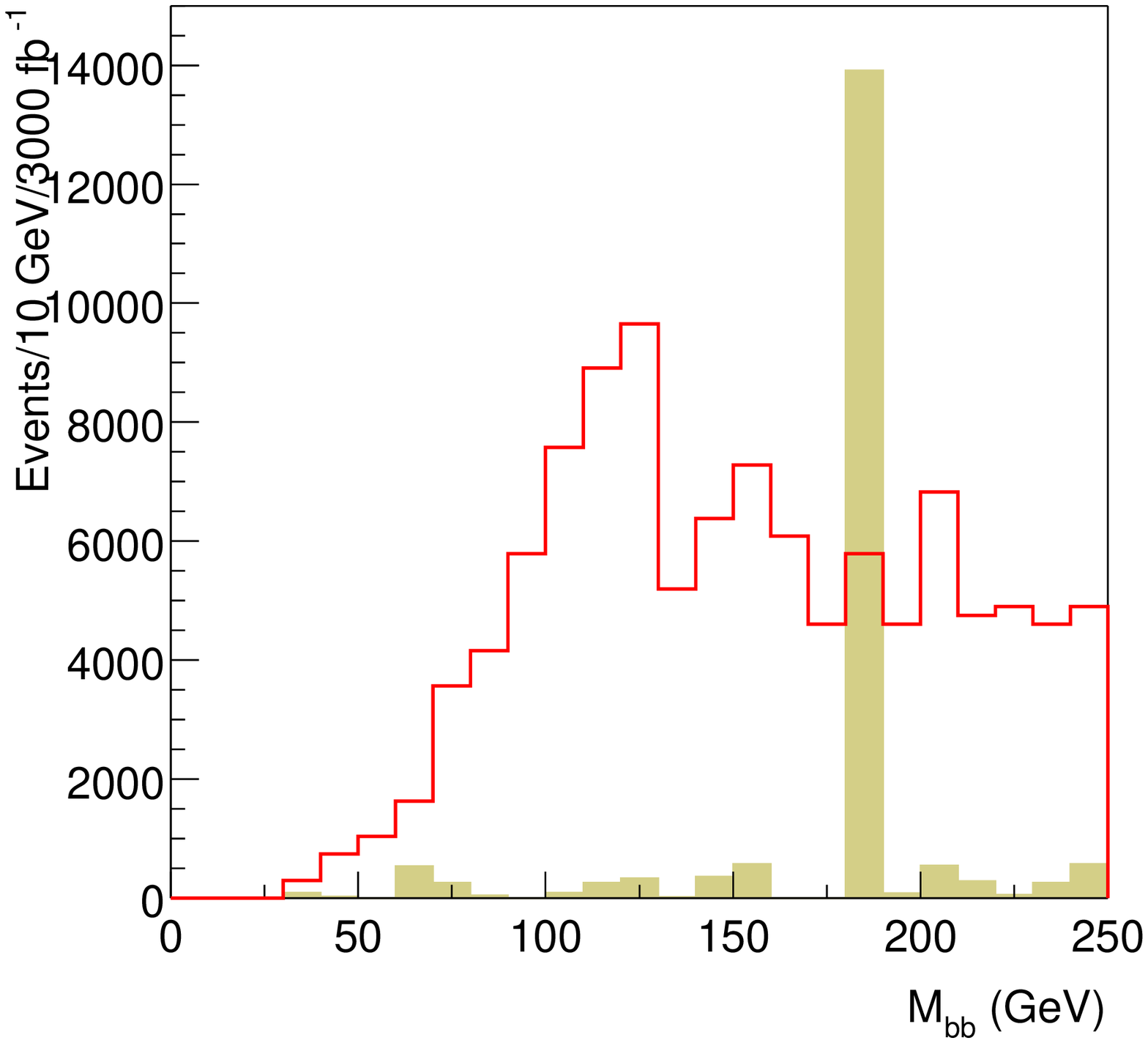}
\caption{$M_{bb}$ distribution for SLHC (left) and VLHC (right) for Point
K. \label{pointkmbb}}
\end{figure}

Production of $\tq_R\tq_R$ followed by the decay of each squark to
$q\lsp$ gives a dijet signal accompanied by missing $E_T$. 
In order to  extract this from the standard model  background, 
hard cuts on the jets and $\etmiss$ are needed.
Events were required
to have two jets with $p_T>700\,\GeV$, $\etmiss>600\,\GeV$, and
$\Delta\phi(j_1,j_2)<0.8\pi$. The resulting distributions are shown in
Figure~\ref{pointkptfroid}. Only a few events survive at the LHC with
3000 $\fbi$. The transverse momentum of the hardest jet is 
 sensitive to the
$\tq_R$ mass\cite{AtlasPhysTDR}. The mass determination will be
limited by the available statistics.

The decay $\tchi_2^0 \to \lsp h$ is dominant so we should expect to
see Higgs particles in the decay of $\tq_L$ ($\tq_L\to \tchi_2^0 q \to
\lsp h q $). The Higgs signal can  be observed as a peak in the
$b\overline{b}$ mass distributions. In order to do this, it is
essential that $b-$jets can be tagged with good efficiency and
excellent 
rejection against light quark jets. 
There is a large
background from $t \bar t$ that must be overcome using topological cuts.
 Events were selected to have at least three
jets with $p_T > 600,300,100\,\GeV$, $\etmiss>400\,\GeV$,
$\Meff>2500\,\GeV$, $\Delta\phi(j_1,\etmiss)<0.9\pi$, and
$\Delta\phi(j_1,j_2)<0.6\pi$. The distributions are shown in
Figure~\ref{pointkmbb} assuming the same $b$-tagging performance as for
standard luminosity, {\it i.e.,}  that shown in  Figure~9-31 of
Ref.~\cite{AtlasPhysTDR} which corresponds to an efficiency of 60\%
and a rejection factor against light quark jets of $\sim 100$.
This $b-$tagging performance may be optimistic in the very high
luminosity environment. However our event 
selection is only $\sim 10\%$ efficient at SLHC and might be improved.
There is much less standard model background at VLHC. However,  there is
significant SUSY background from  $\tg \to \tilde b_i \bar b, \tilde
t_1 \bar t$ which becomes more important at the higher energy.  
At the VLHC and possibly a the SLHC, it should be
possible to extract information on the mass of $\tq_l$ by combining
the Higgs with a jet and probing the decay chain $\tq_L \to  \tchi_2^0
q \to q h \lsp $ (see {e.g.} \cite{precise}).

\section{Point M}

Point M has squark and gluino masses around 3.5 TeV and is beyond the
reach of the standard LHC. Only 375 SUSY events of
all types are produced for $1000\,\fbi$ at LHC, mainly valence squarks
($\tilde u_L, \tilde d_L, \tilde u_R, \tilde d_R$) and gauginos
($\tchi_1^\pm,\tchi_2^0$). The VLHC cross section is a factor of 200
larger. About half of the SLHC SUSY events are from electro weak 
gaugino pair production mostly $\tchi_2^0$ and $\tchi_1^\pm$ . The
dominant decays of these are  $\tchi_2^0 \to \lsp h$ and $\tchi_1^\pm \to \lsp
W^\pm$. Rates are so small that no signal
 close to the Standard Model  backgrounds coul be  found for the
SLHC.

\begin{figure}[t]
\includegraphics[width=3.0in]{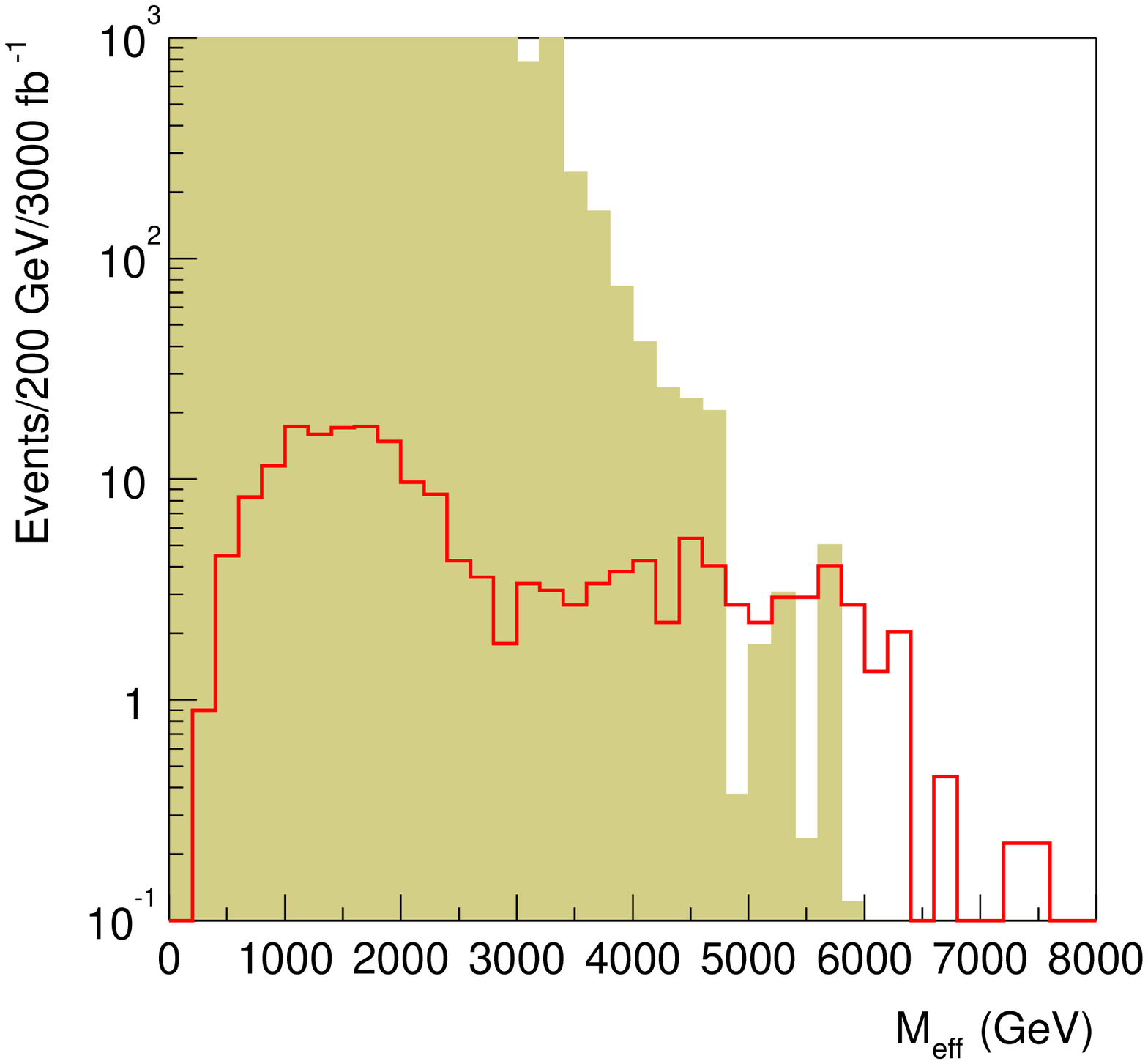}
\includegraphics[width=3.0in]{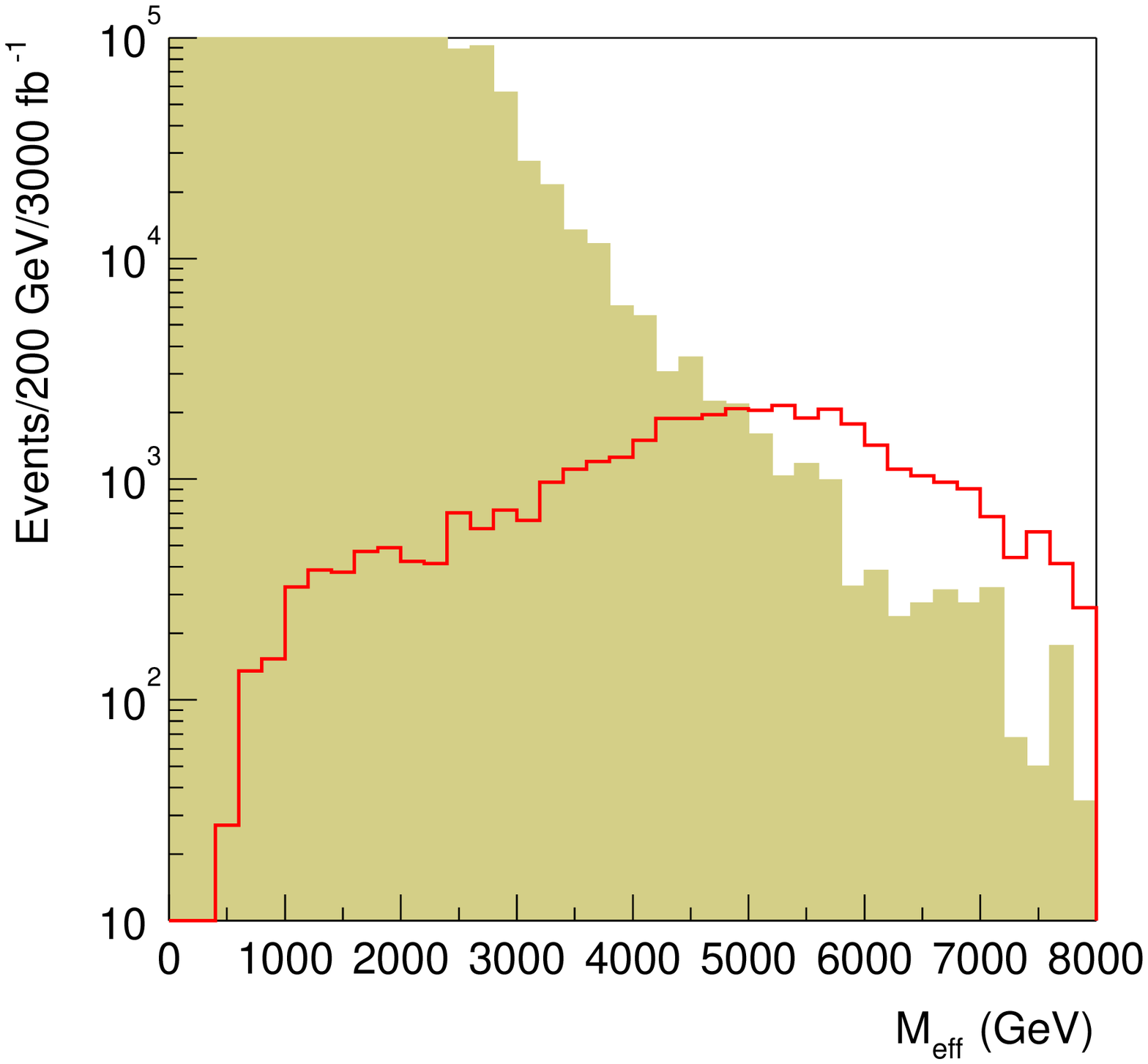}
\caption{$\Meff$ distribution for SLHC (left) and VLHC (right) for
Point M. \label{pointmmeff}}
\end{figure}

\begin{figure}[t]
\includegraphics[width=3.0in]{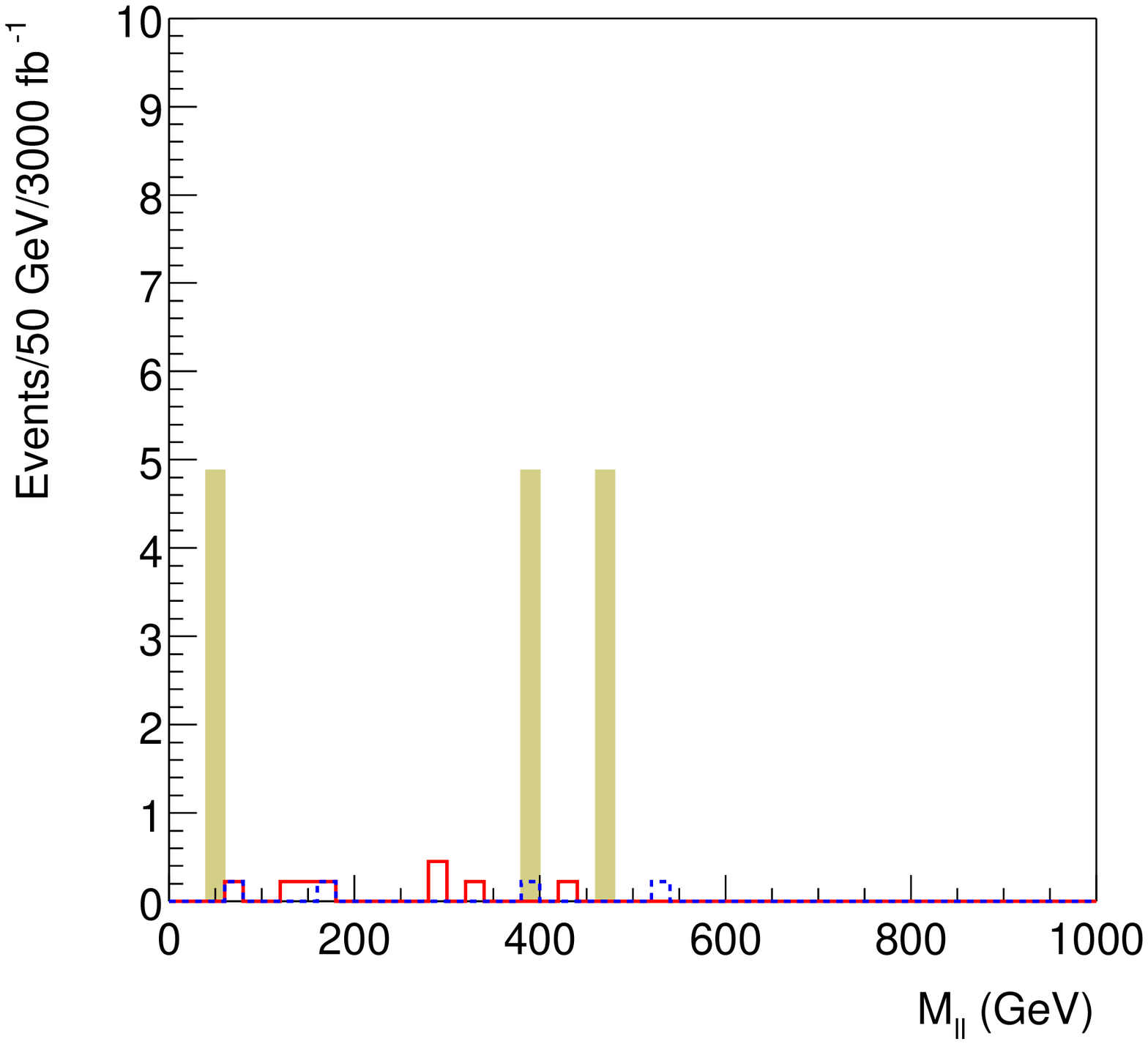}
\includegraphics[width=3.0in]{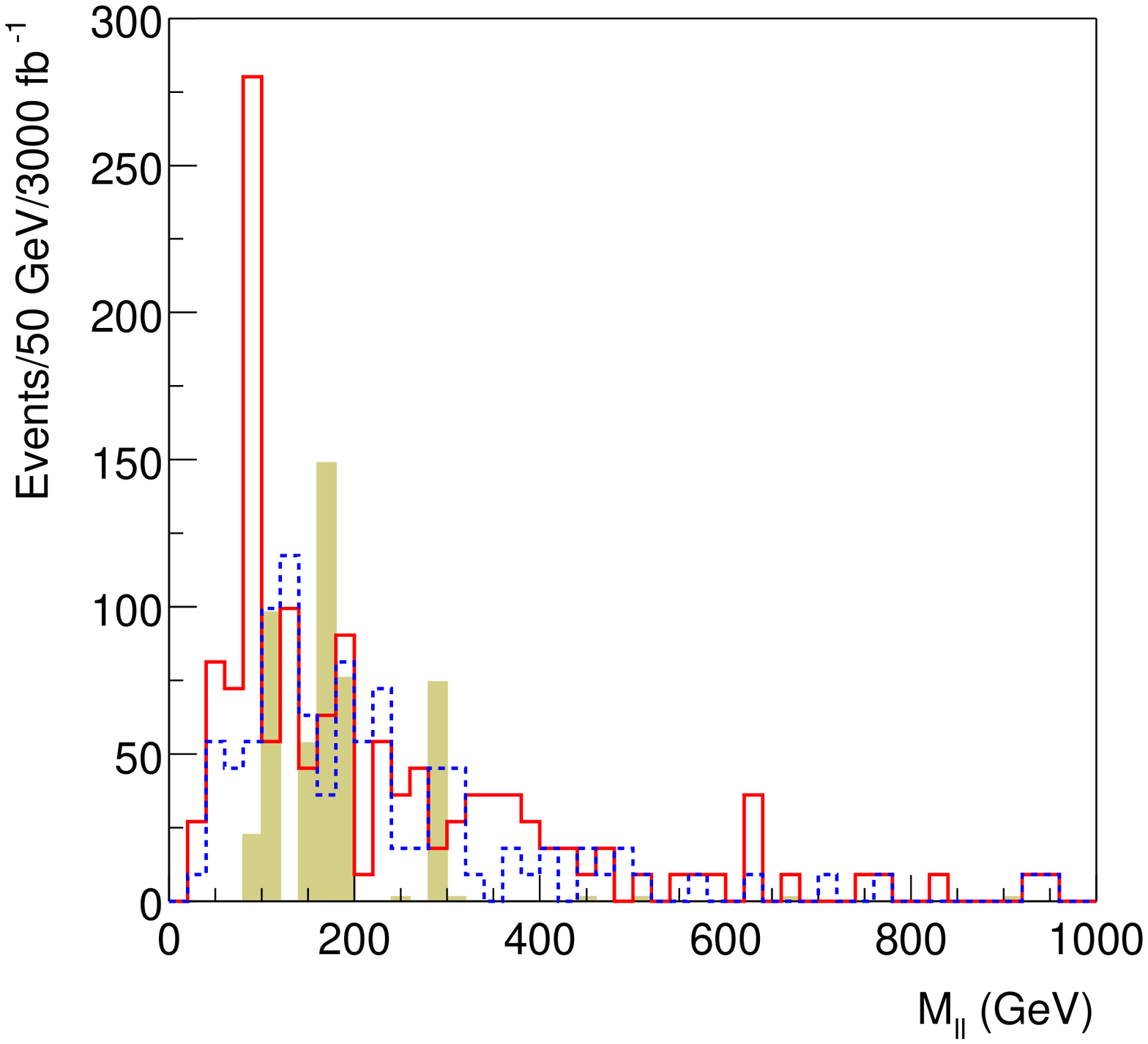}
\caption{Dilepton mass distribution for SLHC (left) and VLHC (right) for
Point M. Solid: $\ell^+\ell^-$. Dashed: $e^\pm\mu^\mp$.
\label{pointmmll}}
\end{figure}

The effective mass distributions for Point M at SLHC and VLHC are shown
in Figure~\ref{pointmmeff} using the same cuts as for Point K. As
expected, the SLHC
signal is very marginal: there are only 20 signal events with 10
background events for $\Meff > 5000\,\GeV$ and $3000 \fbi$. 
Several attempts to optimize
the cuts did not give any improvement. Requiring a lepton, a hadronic
$\tau$, or a tagged $b$ jet  did not help.  We are forced to conclude
than it is unlikely that a signal of any type could be observed.
The VLHC signal is
clearly visible and could be further optimized. 

The dilepton rates are shown in Fig~\ref{pointmmll}. Events are
selected that have $M_{eff}>3000 \GeV$ $\etmiss>0.2M_{eff}$ and two
isolated leptons with $p_T>15 \GeV$ and the mass distribution of the
dilepton pair shown.  As expected, nothing is visible
at SLHC. The distribution at VLHC is dominated by two independent
decays ({\it e.g.} $\tchi_1^\pm \tchi_1^\mp \to \lsp \lsp
W^\mp$),  so that $e^+e^-+\mu^+\mu^-$ and $e^\pm\mu^\mp$ rates are almost
identical except for the $Z$ peak in the former which arises mainly from $\tq
\to q \tchi_2^\pm \to q \tchi_1^\pm Z$.

On the basis of this preliminary, 
 study it seems unlikely that Point M can
be detected at $14\,\TeV$ even with $3000\,\fbi$. Higher energy
would be required.

\section{Point H}

Point H is
able to accommodate very heavy  sparticles  without
overclosing the universe as the destruction rate for the $\lsp$
 is enhanced  by coannihilation  with a stau. This implies a very
small splitting between the $\ttau_1$ and the $\lsp$. In this
particular case,  $\ttau_1
\not\to \lsp \tau$, so it must decay by second order weak processes,
$\ttau_1 \to \lsp e\bar\nu_e\nu_\tau$, giving it a long lifetime. The
dominant SUSY rates arise from the  strong production of  valance squarks, with
$\tq_L \to \tchi_1^\pm q, \tchi_2^0 q$ and $\tilde q_R \to \lsp$. The
stau which are produced from cascade decays of these squarks, then
exit the detector with a signal similar to a ``heavy muon''.

\begin{figure}
\includegraphics[width=3.0in]{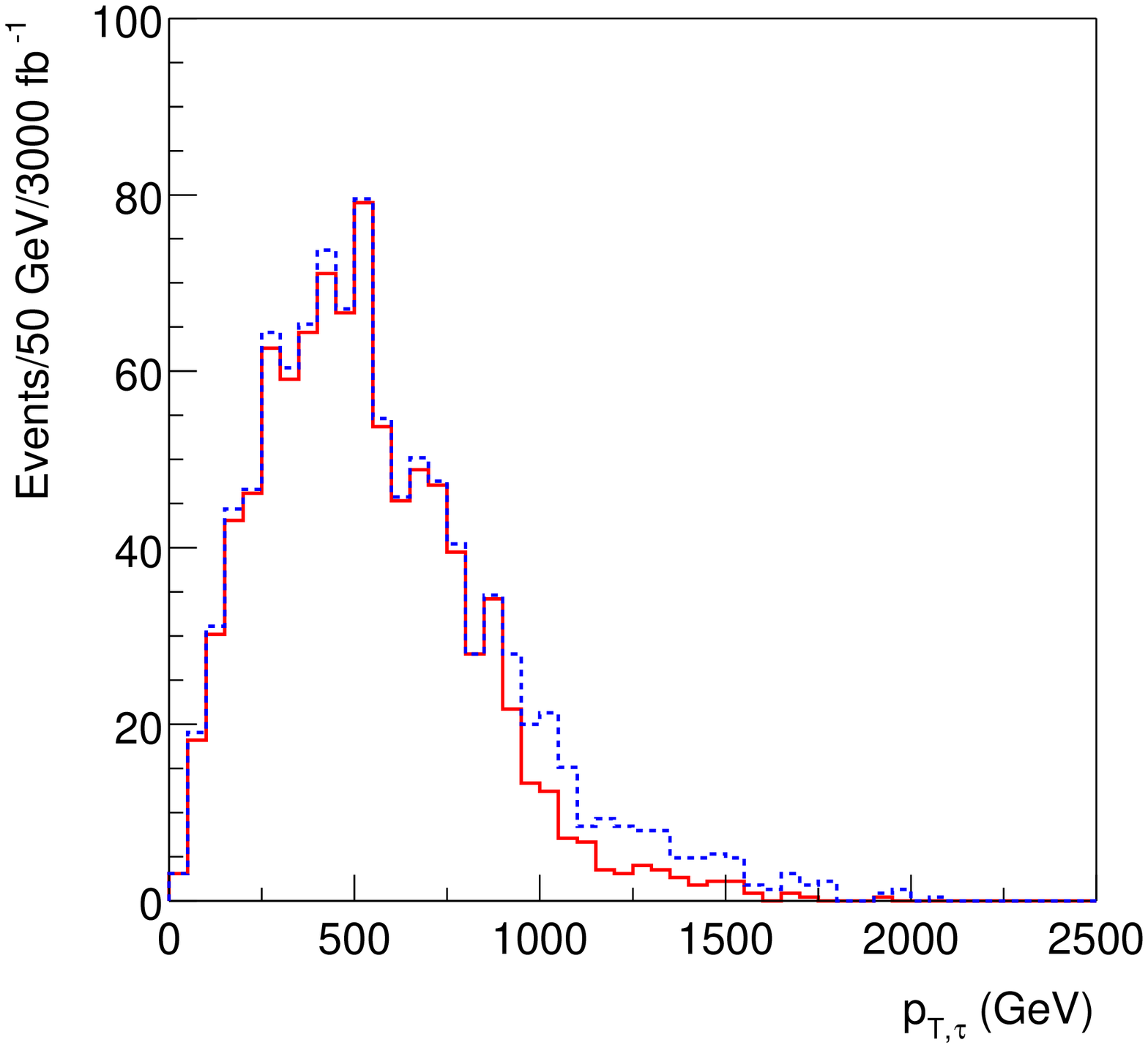}
\includegraphics[width=3.0in]{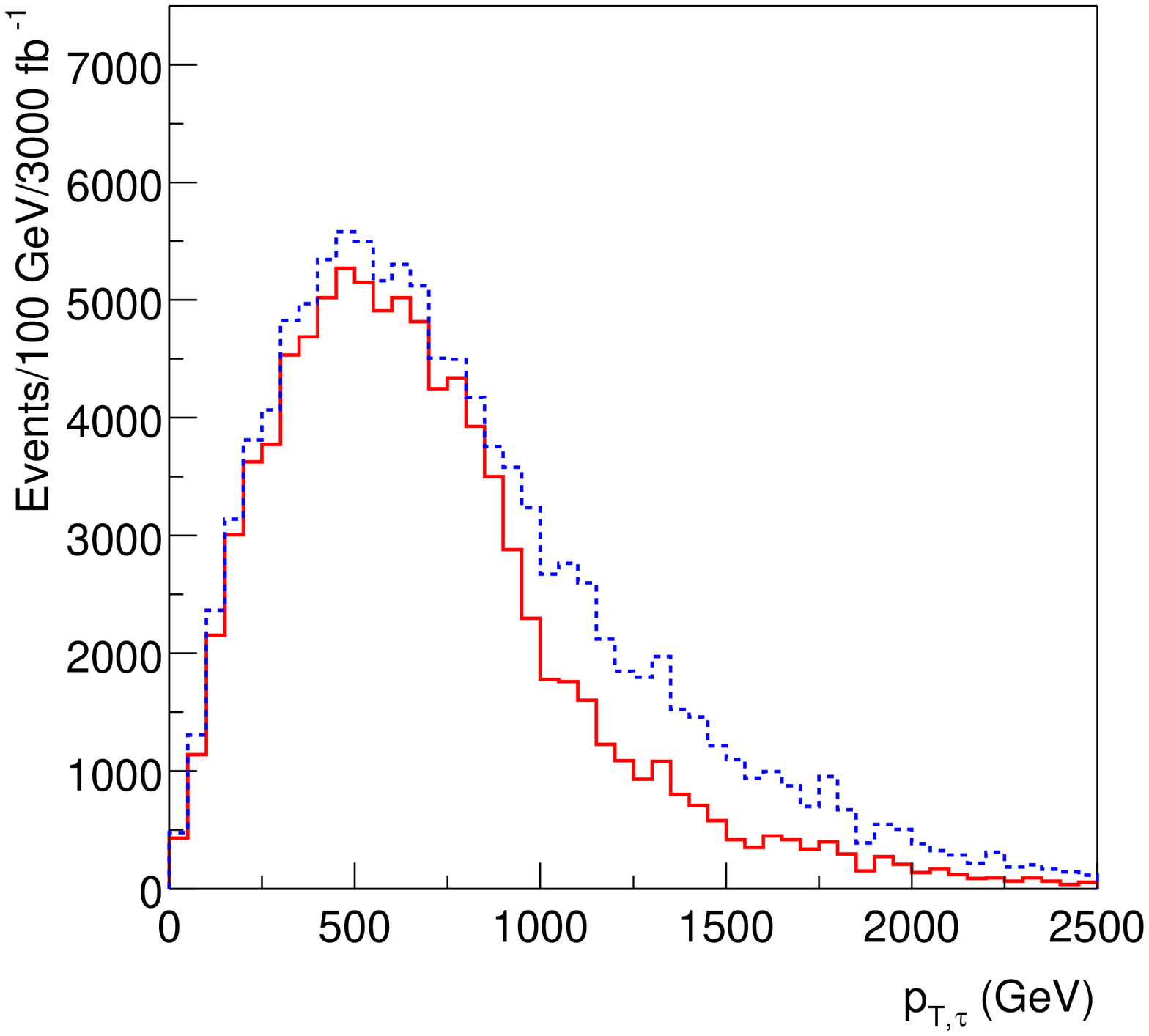}
\caption{$p_T$ distribution of $\ttau_1$ at SLHC (left) and VLHC (right)
for Point H. Dashed: all $\ttau_1$. Solid: $\ttau_1$ with $\Delta t >
7{\,\rm ns}$ \label{pointhptstau}}
\end{figure}

\begin{figure}
\includegraphics[width=3.0in]{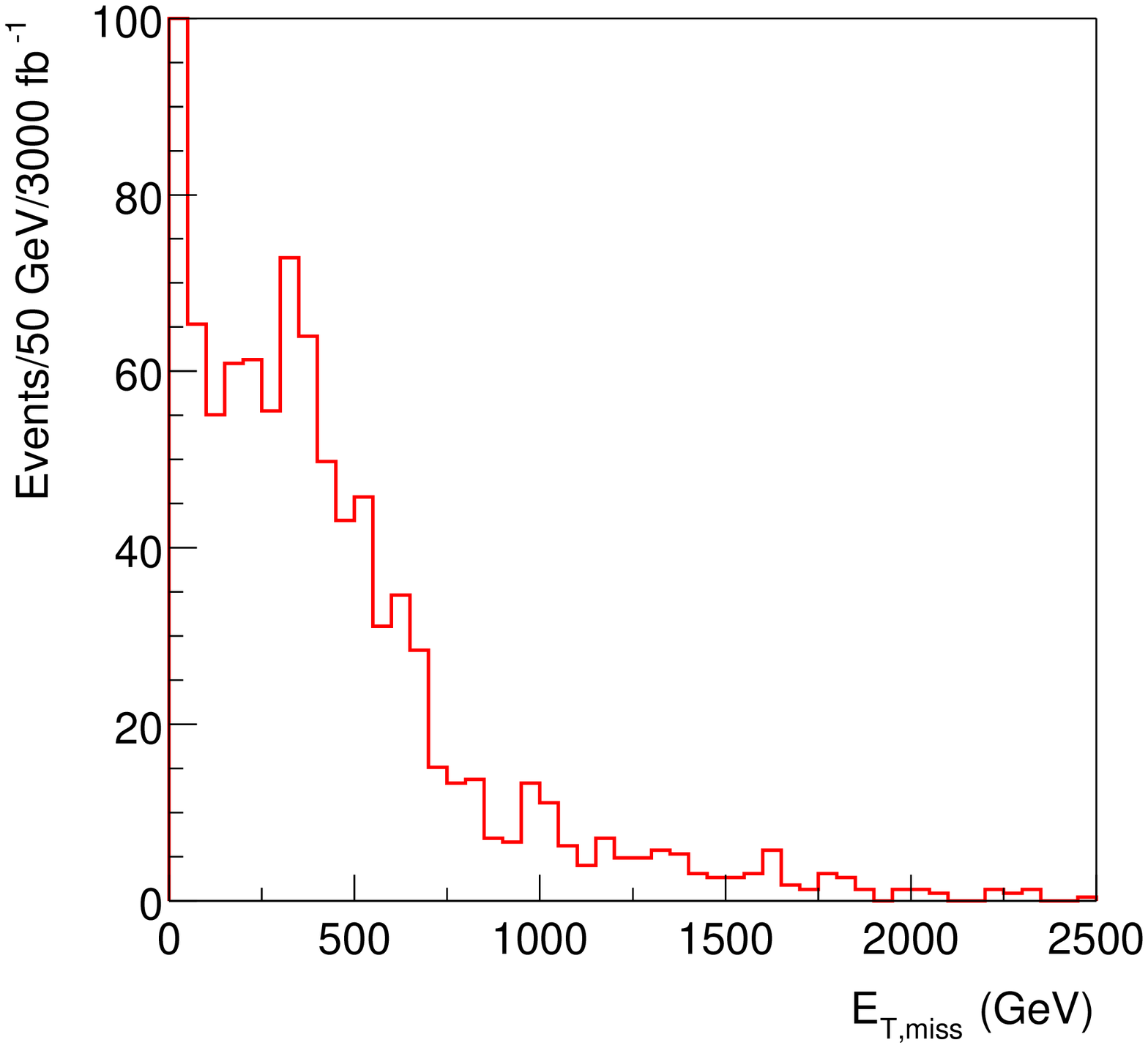}
\includegraphics[width=3.0in]{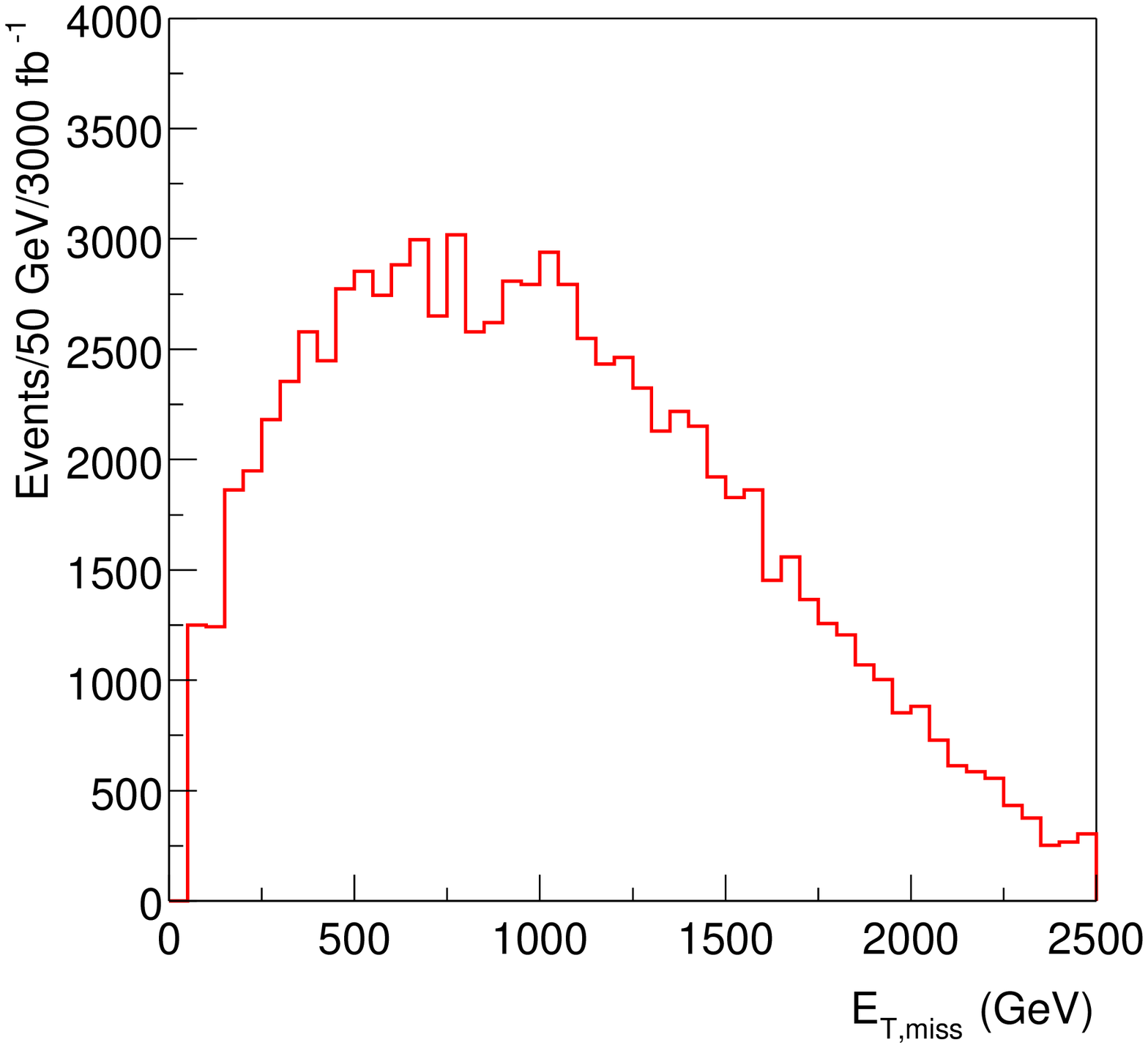}
\caption{Calorimetric $\etmiss$ distributions in $\ttau_1$ events for
SLHC (left) and VLHC (right) for Point H. \label{pointhetxstau}}
\end{figure}

The $p_T$ spectrum of these quasi-stable $\ttau_1$ for $1000\,\fbi$ is
shown in Figure~\ref{pointhptstau}. The ATLAS muon system
\cite{AtlasPhysTDR} has a time
resolution of about 0.7~ns for time of flight over a cylinder of radius
10~m and half-length 20~m. The spectrum with a time delay $\Delta t >
10\sigma(7\,{\rm ns})$ is also shown. Notice that this signal could be
observed at
the LHC with $\sim 300\,\fbi$.
Triggering on a slow $\ttau_1$ may be a problem since the time-window  for the
trigger chambers is limited. 
However, the $\etmiss$ in SUSY events as measured by the calorimeter
is quite large as shown in Figure~\ref{pointhetxstau}.
 It probably is possible to trigger just on
jets plus $\etmiss$, the distribution for which is shown in
Figure~\ref{pointhetxstau}. The mass of the stable stau can be measured
by exploiting the time of flight measurements in the muon measurement
system.
Studies of such quasi stable particles at somewhat smaller masses
carried out or the ATLAS detector showed a mass resolution of
approximately 3\% given sufficient statistics (see Section 20.3.4.2 of
Ref~\cite{AtlasPhysTDR}).
A precision of this order should be achievable with 3000 $\fbi$ at
either the LHC or VLHC. One can then build on the stable stau to
reconstruct the decay chain using techniques similar to those used for
the GMSB studies \cite{AtlasPhysTDR}\cite{Hinchliffe:1999ys}. This is
not pursued here.

\begin{figure}
\includegraphics[width=3.0in]{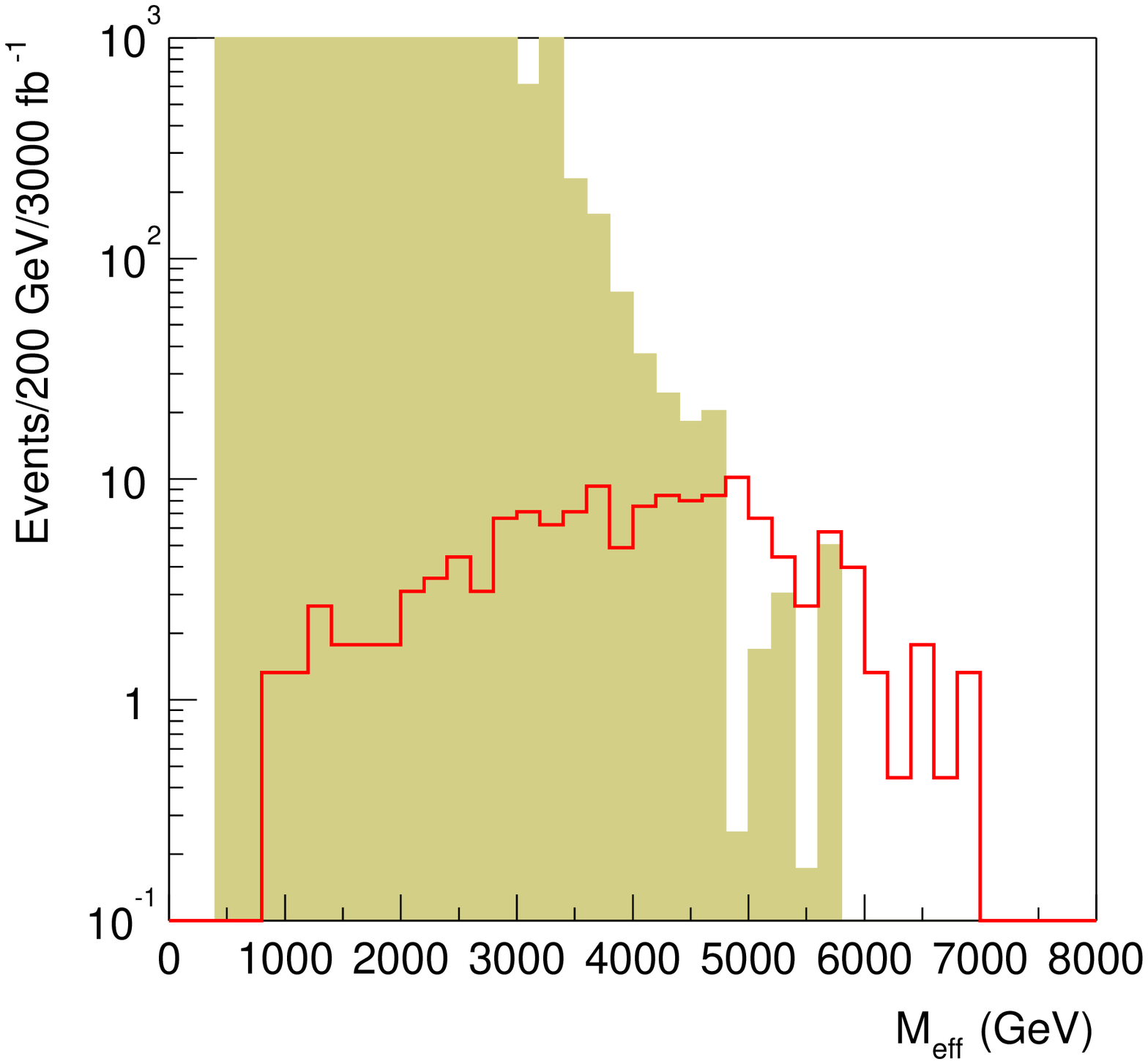}
\includegraphics[width=3.0in]{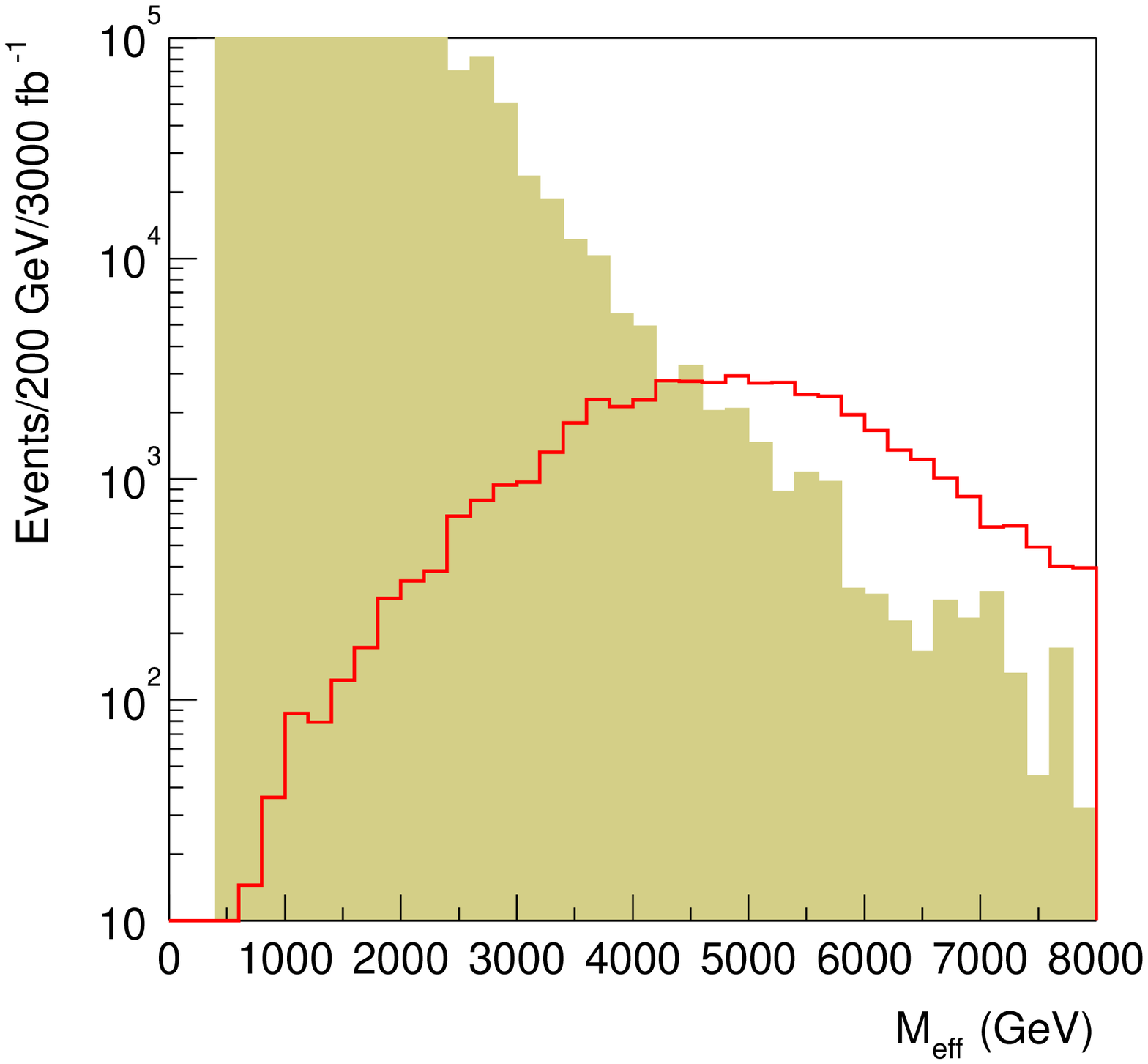}
\caption{$\Meff$ distribution for SLHC (left) and VLHC (right) for Point
H. Solid: signal. Shaded: SM background. \label{pointhmeff}}
\end{figure}

The stable $\tau_1$ signature is somewhat exceptional so we explore
other signatures that do not require it and would be present if the
stau decayed inside the detector. For such high
masses the strong production is mainly of $\tilde u$ and $\tilde d$.
Events are selected with hadronic jets and missing $E_T$ and the
effective mass formed as in the case of Point K.
 To optimize this signature, events were further 
selected with at least two jets
with $p_T > 0.1\Meff$, $\etmiss > 0.3\Meff$, $\Delta\phi(j_0,\etmiss) <
\pi-0.2$, and $\Delta\phi(j_0,j_1)<2\pi/3$. The $\Meff$ distributions
after these cuts for the SLHC and the VLHC are shown in
Fig~\ref{pointhmeff}. Note that at the SLHC the number of events in
the region where $S/B>1$ is very small. 
 Given the uncertainties in the modeling of
the standard model backgrounds the shower Monte Carlo, 
it is not possible to claim that the
SLHC could see a signal using this global variable.
 The VLHC should have no difficulty as there are several thousand
 events for $M_{eff}>5$ TeV.

\begin{figure}
\includegraphics[width=3.0in]{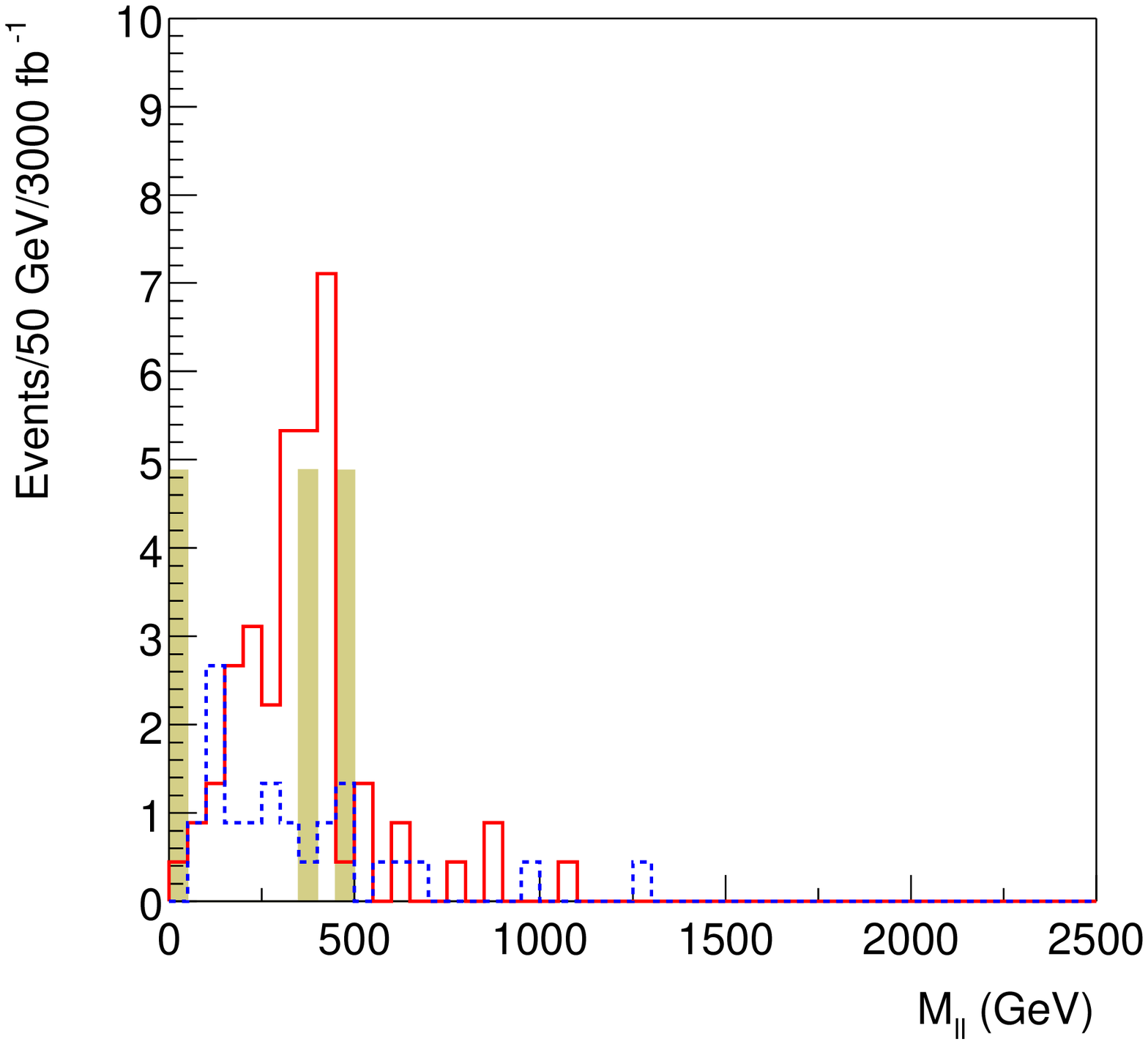}
\includegraphics[width=3.0in]{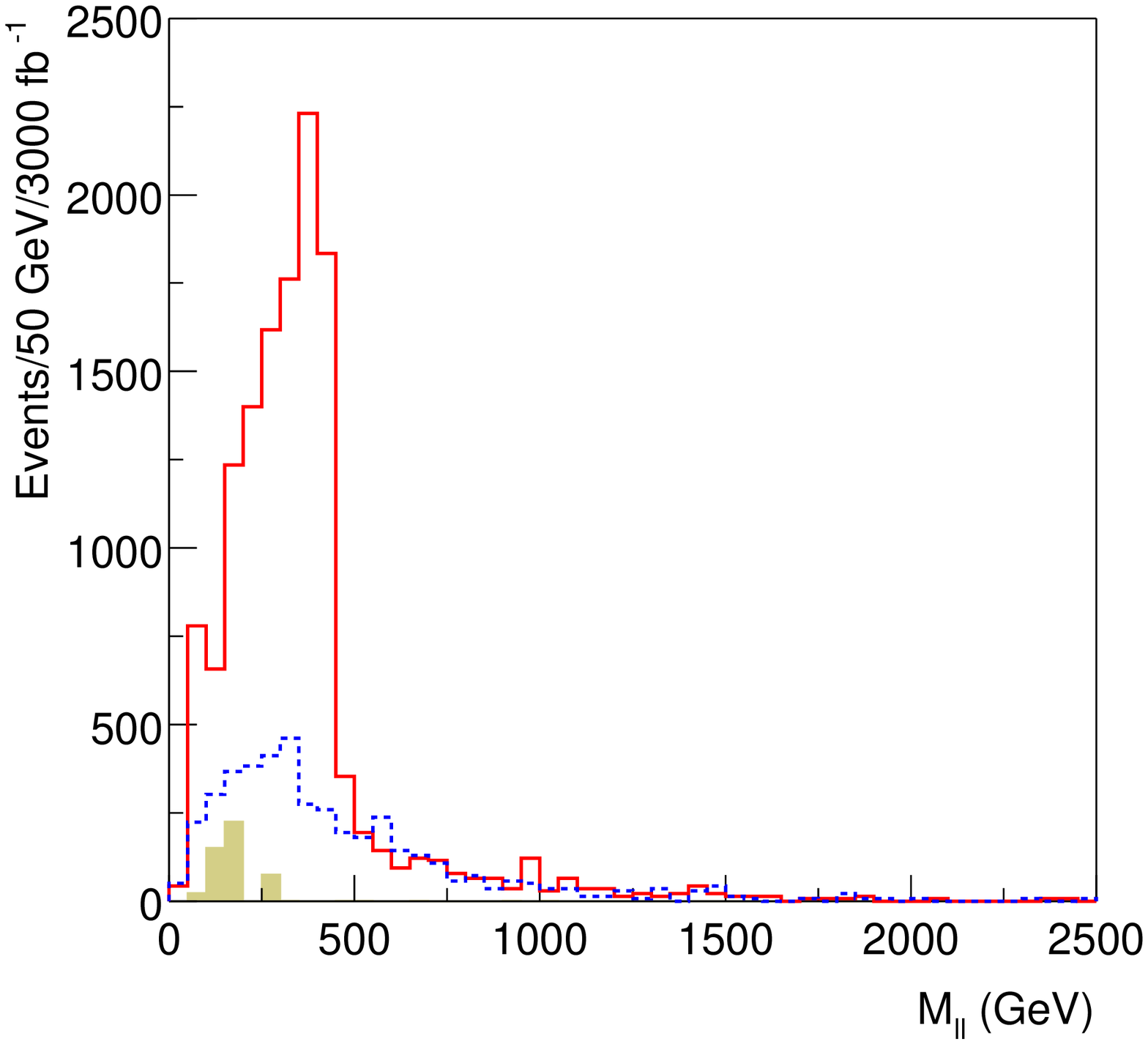}
\caption{$M_{\ell\ell}$ distribution for SLHC (left) and VLHC (right)
for Point H.  Solid: $\ell^+\ell^-$. Dashed: $e^\pm\mu^\mp$.
\label{pointhmll}}
\end{figure}
\begin{figure}
\includegraphics[width=3.0in]{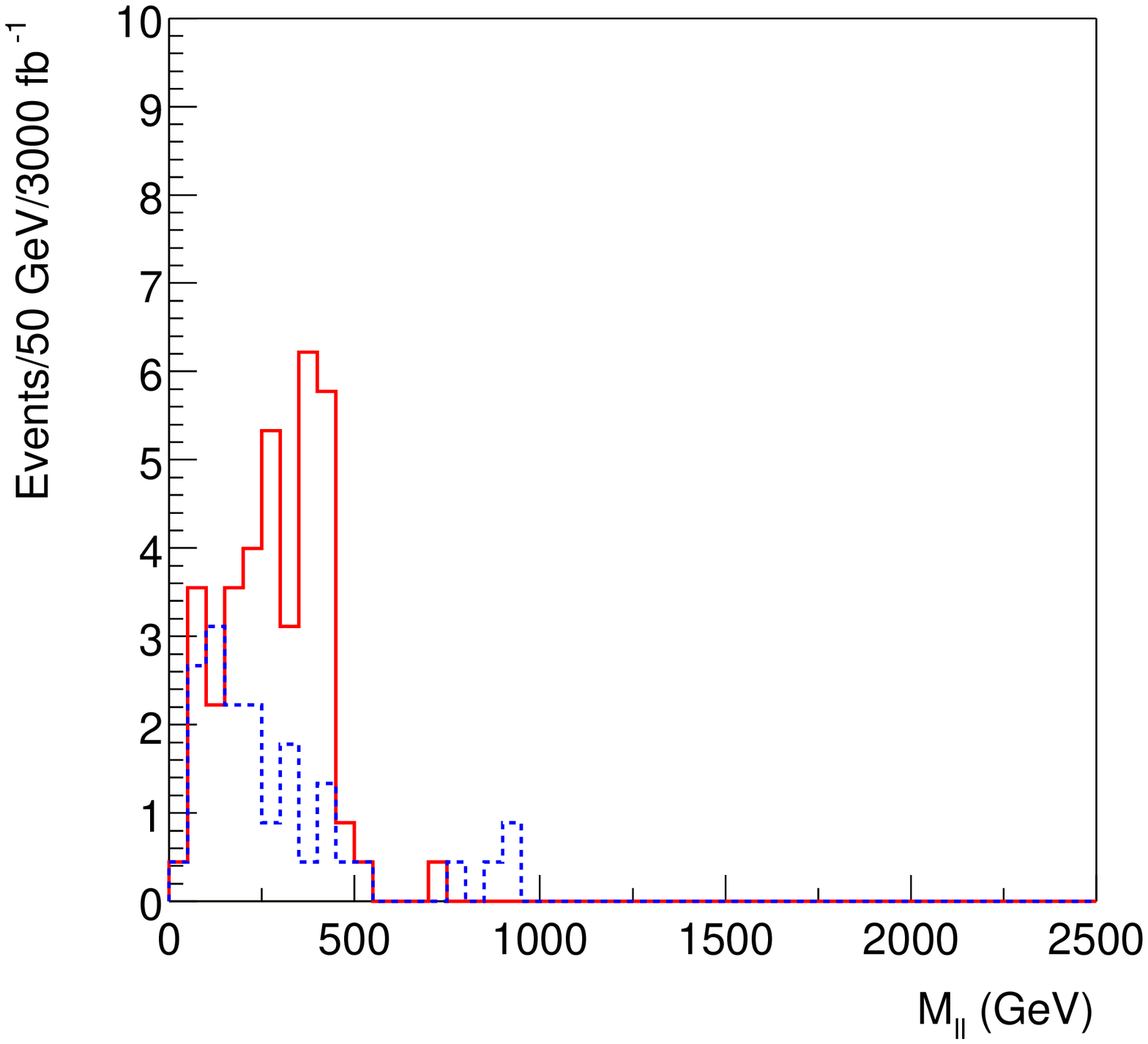}
\includegraphics[width=3.0in]{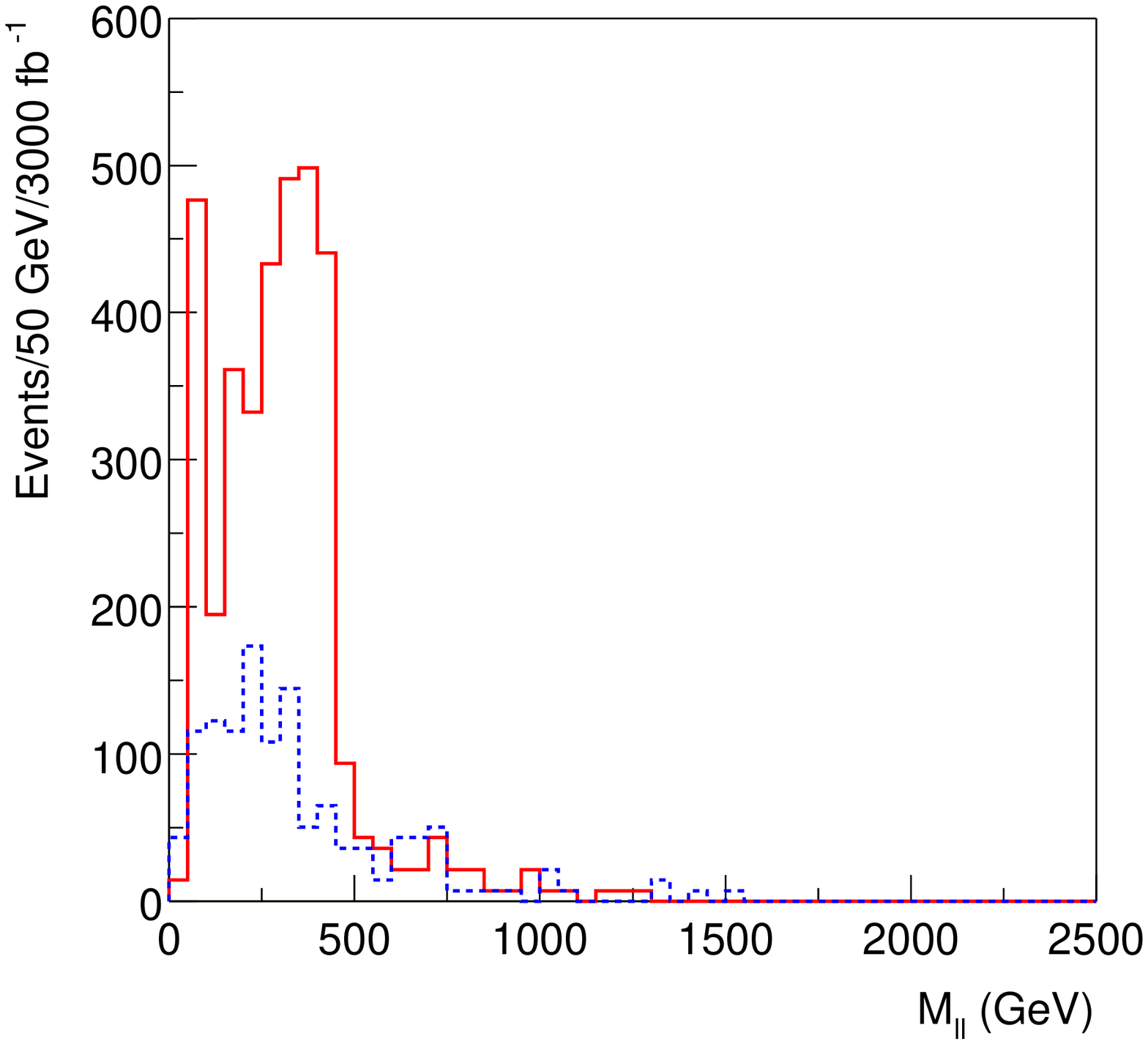}
\caption{$M_{\ell\ell}$ distribution for SLHC (left) and VLHC (right)
for events containing a $\ttau_1$ for Point H.  Solid: $\ell^+\ell^-$.
Dashed:  $e^\pm\mu^\mp$. \label{pointhmllstau}}
\end{figure}

Dileptons arise from the cascade $\tq_L\to q \tchi_2^0\to
q\ell^+\ell^- \lsp$, 
The dilepton mass distributions should have a kinematic endpoint
corresponding to this decay. Figure~\ref{pointhmll} shows the
distribution for same flavor and different flavor lepton pairs.
Events were required to have $\Meff>3000\,\GeV$
and $\etmiss>0.2\Meff$ and to have two isolated opposite sign leptons
with $E_T>15$ GeV and $\abs{\eta}<2.5$.
 The structure at the VLHC is clear;
the edge comes mainly from $\tchi_2^0 \to
\tell_L^\pm \ell^\mp$, which has  a branching ratio of 15\% per flavor. This
gives an endpoint at
$$
\sqrt{(M_{\tchi_2^0}^2-M_{\tell_L}^2)(M_{\tell_L}^2-M_{\lsp}^2)
\over M_{\tell_L}^2} = 447.3\,\GeV
$$
consistent with the observed endpoint in Figure~\ref{pointhmll}. Of course
this plot does not distinguish $\tell_L$ and $\tell_R$. In the case of
the upgraded LHC, the signal may be observable, but it should be noted
that the background is uncertain as only three generated events passed
the cuts.

If the stable stau is used then the situation improves considerably.
The dilepton mass for events containing a $\ttau_1$ with a time delay $7
< \Delta t < 21.5\,{\rm ns}$ is shown in Figure~\ref{pointhmllstau}.
Since $\Delta t>10\sigma$, the standard model  background is expected  to be
negligible. The SLHC signal is  improved and a measurement should   be possible.
The acceptance for VLHC is somewhat worse than the inclusive sample, but
having the correlation of the dileptons with the $\ttau_1$ should be
useful.

The VLHC gives a gain of $\sim 100$ in statistics over the LHC for the
same luminosity at this point, which is at the limit of observability at
the LHC. If the VLHC luminosity were substantially lower, the
improvement provided by it would be rather marginal. The cross section
increases by another factor of $\sim100$ at $200\,\TeV$.

\section{Conclusions}

We have surveyed the signals at hadron colliders for the SUGRA models
proposed by \cite{Battaglia:2001zp} concentrating on the cases where
the sparticle masses are very large.
While the masses of the sparticles at Point K are such that SUSY would
be discovered at the baseline LHC, the event rates are small and
detailed SUSY studies will not be possible.
The reach of the LHC for  would be improved by
higher luminosity where the extraction of specific final states will
become possible.  The cross section at a 
$40\,\TeV$ VLHC is approximately $100$ times larger than that at LHC.
 This leads to a substantial gain, but it is important to
emphasize that this gain 
requires  luminosity at least as large as that ultimately reached by
the LHC
and detectors capable of exploiting it. Point H has a special feature
in that the stau is quasi-stable. This feature would enable a signal
to be extracted at SLHC. If the tau mass were raised slightly so that
its lifetime were short, then only the VLHC could observe it. The
masses in the case of Point M are so large that the VLHC would be
required for discovery. Point F has a gluino mass of order 2 TeV and
should be observable at the LHC exploiting the production of gluinos
followed by the decays to $\tchi_i$ and hence to leptons.

The Points A, B, C, D,  G, I, and L  which are much less fine tuned
have  similar phenomenology to  the ``Point 5'' or
``Point 6'' 
analysis of \cite{AtlasPhysTDR} in that lepton structure from the
decay $\tchi_2^0\to \tell_R \ell \to \ell^+\ell^- \lsp$ and/or
$\tchi_2^0\to \tilde{tau}\tau \to \tau^+\tau^- \lsp$ is present.
In most cases
decay
$\tchi_2^0\to \tell_L\ell$ is also allowed, so that a more complicated
dilepton mass spectrum is observable. This should enable the
extraction of $m_{\ell_L}$ in addition
(for an example see Fig 20-53 of \cite{AtlasPhysTDR}). Points A, D
and L have higher squark/gluino masses and will require more
integrated luminosity. Nevertheless one can have confidence that the
baseline LHC will make many measurements in all of these cases.

The work was supported in part by the Director, Office of Energy
Research, Office of High Energy and Nuclear Physics of the U.S. Department of Energy under Contracts
DE--AC03--76SF00098 and DE-AC02-98CH10886.  Accordingly, the U.S.
Government retains a nonexclusive, royalty-free license to publish or
reproduce the published form of this contribution, or allow others to
do so, for U.S. Government purposes.

\end{document}